\documentstyle[12pt]{article}

\begin{document}
\begin{center}
{\Large \bf On the decay of neutral kaons }
\bigskip

{\large D.L.~Khokhlov}
\smallskip

{\it Sumy State University, R.-Korsakov St. 2, \\
Sumy 244007, Ukraine\\
E-mail: khokhlov@cafe.sumy.ua}
\end{center}

\begin{abstract}
It is shown that the data on the decay of neutral kaons
may be explained without CP-violation.

\end{abstract}

As known~\cite{Co} strange $K^0$ and $\bar K^0$ mesons do not
posess surtain lifetimes relative to the weak decays, since
weak interactions do not conserve the strangeness number.
There exist two
independent linear combinations of $K^0$ and $\bar K^0$
\begin{equation}
K^0_S=K^0 - \bar K^0
\label{eq:KS}
\end{equation}
\begin{equation}
K^0_L=K^0 + \bar K^0
\label{eq:KL}
\end{equation}
They correspond to the particles with the lifetimes
$\tau_S=8.9\times 10^{-11}\ {\rm s} $ and
$\tau_L=5.2\times 10^{-8}\ {\rm s} $ respectively.
The states $K^0_S$ and $K^0_L$ are
of CP-invariance with the eigenvalues +1 and -1 respectively.
$K^0_S$ decays into the system of two pions
\begin{equation}
K^0_S \rightarrow \pi\pi
\label{eq:KSd}
\end{equation}
with the CP-eigenvalue +1,
and $K^0_L$
decays into the system of three pions
\begin{equation}
K^0_L \rightarrow \pi\pi\pi
\label{eq:KLd}
\end{equation}
with the CP-eigenvalue -1.

$K^0$ is considered as a superposition of
$K^0_S$ and $K^0_L$
\begin{equation}
|K^0>=\frac{1}{\sqrt 2}(K^0_S + K^0_L).
\label{eq:K0}
\end{equation}
Evolution of $K^0$ is given by
\begin{equation}
|K^0(t)>=\frac{1}{\sqrt 2}(K^0_S e^{-t/2\tau_S} +
K^0_L e^{-t/2\tau_L}).
\label{eq:K0t}
\end{equation}
One can expect that, within the time $t<\tau_S$, $K^0$ decays
into two pions, and within the time $\tau_S<t<\tau_L$, $K^0$ decays
into three pions. But, within the time $\tau_S<t<\tau_L$,
there exists the probability of the decays
of $K^0$ into two pions
\begin{equation}
\frac{\Gamma(K^0(\tau_S<t<\tau_L)\rightarrow \pi^+\pi^-)}
{\Gamma(K^0(\tau_S<t<\tau_L)\rightarrow all)}
\approx 2\times 10^{-3}
\label{eq:GG}
\end{equation}
\begin{equation}
\frac{\Gamma(K^0(\tau_S<t<\tau_L)\rightarrow \pi^0\pi^0)}
{\Gamma(K^0(\tau_S<t<\tau_L)\rightarrow all)}
\approx 10^{-3}.
\label{eq:GG0}
\end{equation}
Decays $K^0\rightarrow \pi\pi$ within the time $\tau_S<t<\tau_L$
are treated as CP-violation.

$K^0$ decays in combination with $\bar K^0$.
The state of $K^0$ is defined by the probabilities of decay
of $K^0$ in combinations with $+\bar K^0$ and $-\bar K^0$.
The state of $K^0$ embedded in the $K^0 \bar K^0$ vacuum
is given by
\begin{equation}
|K^0>=\left(1-\frac{\tau_S}{\tau_L}\right)^{1/2} K^0_S +
\left(\frac{\tau_S}{\tau_L}\right)^{1/2} K^0_L.
\label{eq:K00}
\end{equation}
In view of eqs.~(\ref{eq:KS}), (\ref{eq:KL}),
the decay of $K^0_S$ leads to the birth of $K^0_L$
\begin{equation}
K^0_L=K^0 - K^0_S,
\label{eq:KL1}
\end{equation}
and the decay of $K^0_L$ leads to the birth of $K^0_S$
\begin{equation}
K^0_S=K^0 - K^0_L.
\label{eq:KS1}
\end{equation}
Hence evolution of $K^0$ is given by
\begin{eqnarray}
|K^0(t)> & =\displaystyle\frac{1}{\sqrt 2} & \left\{
\left(1-2\frac{\tau_S}{\tau_L}\right)^{1/2}
\left[K^0_S e^{-t/2\tau_S} + (1- e^{-t/\tau_S})^{1/2} K^0_L
e^{-t/2\tau_L}\right] + \right.\nonumber \\
& & \left.\left(2\frac{\tau_S}{\tau_L}\right)^{1/2}
\left[K^0_L e^{-t/2\tau_L} + (1- e^{-t/\tau_L})^{1/2} K^0_S
e^{-t/2\tau_S}\right]\right\}.
\label{eq:K0t1}
\end{eqnarray}

In view of eq.~(\ref{eq:K0t1}), within the time $t<\tau_S$,
the number of $K^0$ decayed into two pions is estimated as
\begin{equation}
\frac{\Gamma(K^0(t<\tau_S)\rightarrow \pi\pi)}
{\Gamma(K^0(t<\tau_L)\rightarrow all)}=
\frac{1}{2}-\frac{\tau_S}{\tau_L}.
\label{eq:fr2}
\end{equation}
Within the time $\tau_S<t<\tau_L$,
the number of $K^0$ decayed into two pions is estimated as
\begin{equation}
\frac{\Gamma(K^0(\tau_S<t<\tau_L)\rightarrow \pi\pi)}
{\Gamma(K^0(\tau_S<t<\tau_L)\rightarrow all)}
=2\frac{\tau_S}{\tau_L}=3.4\times 10^{-3}.
\label{eq:G}
\end{equation}
The above consideration allows to explain data on the decay of $K^0$
without CP-violation.

\end{document}